\documentclass[%
 reprint,
 amsmath,amssymb,
 aps,
]{revtex4-2}

\usepackage{graphicx}
\usepackage{dcolumn}
\usepackage{bm}
\usepackage{siunitx}
\usepackage{booktabs}

\usepackage{multirow}

\begin{document}

\preprint{APS/123-QED}

\title{Multiparameter estimation for the superresolution of two incoherent sources}

\author{Antonin Grateau$^{1}$}
\author{Alexander Boeschoten$^{1}$}
\author{Tanguy Favin-Lévêque$^{1}$}
\author{Isael Herrera$^{1}$}
\author{Nicolas Treps$^{1}$}
\email{nicolas.treps@lkb.upmc.fr}

\affiliation{$^{1}$Laboratoire Kastler Brossel, Sorbonne Université, ENS-Université PSL, CNRS, Collège de France, 4 Place Jussieu, F-75252 Paris, France}


\date{\today}

\begin{abstract}

We experimentally demonstrate the simultaneous estimation of the three parameters characterizing a pair of incoherent optical sources in the sub-Rayleigh regime, enabling super-resolved scene characterization. Using spatial-mode demultiplexing (SPADE) with two demultiplexers—one deliberately shifted—we determine separations well below the diffraction limit and achieve sensitive joint estimation of separation, centroid, and relative brightness over a broad range of scene configurations in a single experimental setting. We benchmark our performance using Fisher-information-based Cramér–Rao bounds, and discuss the corresponding quantum limits. We investigate two complementary scenarios: a realistic case with slightly non-identical sources, and an idealized case of indistinguishable sources.

\end{abstract}

\maketitle


\section{\label{sec:level1}Introduction}

Imaging is the process of extracting information about a system or object from the light that it emits or scatters. It can therefore be formulated as a parameter estimation problem: one or more parameters -- for example, the object's position, its size, and its brightness -- are encoded in the light, the carrier of information, and need to be estimated \cite{helstrom1969quantum, Hol73,Hel1973,braunstein1994statistical, barbieri2022, giovannetti2011advances,gessner2023estimation,Sorelli_2024}. The detection of the light and estimation strategy of the parameter from the data constitute the measurement and determine the quality of the estimation and its fundamental limitations. 

Conventionally, imaging employs a pixelized intensity measurement, also known as direct imaging (DI). However, for certain imaging tasks, DI has severe limitations and other measurement strategies should be employed. An important example is the estimation of the separation between two incoherent point sources, a single-parameter estimation problem. DI is limited by diffraction of the light on a finite aperture, and therefore unable to resolve objects when their overlap is considerably larger than the size of the point-spread function \cite{rayleigh1879xxxi,abbe1873beitrage,sparrow1916spectroscopic, shahram2006statistical}. Evaluating this problem from a quantum metrology point of view, it becomes clear that this is not a fundamental limitation. In particular, Tsang \textit{et al.} \cite{TsaNaiLu2016} showed that whereas in DI the information per photon vanishes at small separations, the quantum Cramér--Rao bound (QCRB), which sets the ultimate precision limit, remains finite and separation independent, even in the limit of zero separation. It was shown that an intensity measurement in the Hermite–Gauss (HG) mode basis, known as spatial mode demultiplexing (SPADE), achieves this bound.

This technique has been experimentally implemented, and the superiority of SPADE over DI is verified in several experiments using different techniques \cite{Rouviere:24,Paur:16,Tang:16,Yang:16,Tham:17,Zhou:19,Grenapin:23,SanSgoLup2024}. However, realistic imaging scenarios are inherently multiparametric. For instance, the image of two point sources is defined by their separation, their centroid position, and their relative intensities. These parameters are simultaneously encoded in the optical field and are not independent \cite{rehacek2017multiparameter,Rehacek2018}. Therefore, the ultimate precision limits and the optimal measurement strategies require the framework of multiparameter estimation theory \cite{liu2020quantum, DemGorGut2020, boeschoten2025estimation}. From this perspective, multiparameter estimation can be viewed as the step between single-parameter estimation and full image reconstruction, thereby enabling super-resolved imaging of simple scenes \cite{Delaubert:06, Grace:20,grace2022, Guha2023, Guha2024, Frank:23, duplinskiy2025tsang}.

The joint estimation of centroid and distance has been performed with different techniques, all based on SPADE \cite{TanCheTsa2023,ParBorDem2018}. However, these experiments still assume the sources to have equal intensity. Recently, the simultaneous estimation of the intensity imbalance and separation, assuming a known barycenter, was performed \cite{wallis2025spatial}, but the full three parameter estimation has only been done in the time domain and sequentially \cite{AnsBreSilb2021}. Indeed, the real challenge of multiparameter estimation lies not only in achieving high precision for a specific scene, but also in designing a measurement strategy that provides robust sensitivity over a wide range of possible scenes, without prior knowledge of the parameters to be estimated.

In this letter, we experimentally perform the simultaneous estimation of the separation, centroid, and intensity imbalance of two incoherent point sources over a broad range of parameter values. This is done in a single experimental setting, without adapting the measurement to the unknown scene. Our approach relies on two spatial demultiplexers (Multi-Plane Light Converters, MPLCs), with one deliberately shifted to enhance the information. This scheme is more robust and versatile than implementing the full set of optimal modes \cite{Rehacek2018,AnsBreSilb2021}, which are at the same time scene-dependent and challenging to realize in practice, and thus experimentally relevant. Importantly, our analysis highlights the subtleties that arise when moving from the ideal theoretical case to realistic experimental conditions, where the two sources are slightly different, and paves the way for practical applications.

In what follows, we first show how the dual-MPLC configuration enhances parameter sensitivity compared to a single-MPLC setup. We then present results for experimental source configurations, and finally address the more fundamentally challenging scenario of indistinguishable sources, investigated using data obtained from a single light source.

\section{Two mode-basis approach for multiparameter estimation}

Performing multiparameter estimation is not only about achieving good precision for a specific scene, but also about designing an experimental configuration that provides robust sensitivity over a wide range of possible scenes. In this work, rather than tailoring an optimal mode decomposition for a particular parameter set, we aim at a versatile measurement strategy that remains informative across the full parameter space. To this end, we introduce a dual-basis approach based on two commercial multi-plane light converters (MPLCs), which decompose the incoming field into two shifted Hermite--Gaussian mode bases.

It is well known that a single Hermite--Gaussian basis is insufficient to simultaneously estimate the source separation $d$, centroid $c$, and brightness imbalance $p$, due to the intrinsic symmetries of this basis. Previous work has shown that the unavoidable asymmetries of a practical MPLC can be exploited to access limited multiparameter information for equally bright sources \cite{TanCheTsa2023}. However, this effect is weak and does not enable three-parameter estimation. We therefore use two mode bases: the first MPLC provides a practical Hermite--Gaussian mode decomposition, while the second MPLC is deliberately shifted laterally, resulting in a complementary mode basis. In practice, we use the first four horizontal modes of the first MPLC to extract most of the available information, and we include a single mode from the shifted MPLC to lift remaining degeneracies. This addition significantly improves robustness, reducing estimator bias and enhancing sensitivity across the parameter space.

To quantify the benefit of the dual-MPLC configuration, we compare its theoretical sensitivity to that of a single-MPLC setup using experimentally realistic conditions, including imperfect mode decomposition and source non-idealities. We construct estimators directly from experimentally measured calibration curves. By using either one or two sets of calibration curves, we can model distinguishable and indistinguishable sources, respectively. For each configuration, we compute the classical Fisher information matrix and the associated Cramér--Rao bounds (CRBs), assuming shot-noise-limited detection (see Supp. Mat.). These bounds provide a lower limit on the achievable standard deviation for each parameter $\alpha \in \{d,c,p\}$,

\begin{equation}
\sigma_{\alpha}\geq \sigma_{\alpha,\mathrm{CRB}}=\sqrt{(\mathcal{F}^{-1})_{\alpha\alpha}}. 
\end{equation}

We perform our simulations on an ensemble of scenes spanning the relevant parameter range at fixed total photon number. For each configuration (one versus two MPLCs) and for each scene, we compute the full Fisher information matrix and extract the marginal Cramér--Rao bounds on each parameter. 

\begin{figure}[h]
\includegraphics[width=1\linewidth]{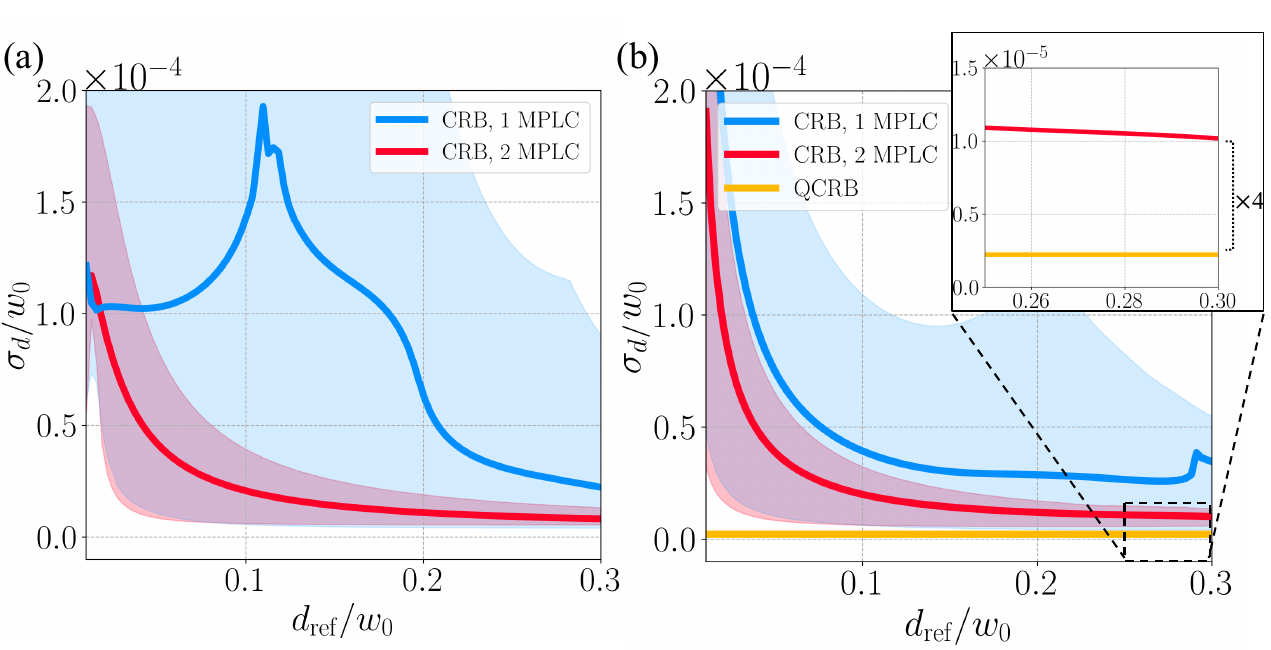}
\caption{\label{fig:NumericalSensitivity} Cramér--Rao bounds associated with $d$, for one and two MPLC configuration associated with (a) distinguishable and (b) indistinguishable sources. In solid lines we show the average precision on $d$. The corresponding shaded area is the region containing $90\%$ of the computed values. We use an ensemble of scenes defined by $d\in [0.01,0.3]w_0$, $c\in [-0.15,0.15]w_0$  and $p= 0.5$, with $N=10^{11}$ signal photons. For indistinguishable sources we also show the quantum CRB, which exhibits nearly constant value of $\approx 2.3\times10^{-6}$.  %
}
\end{figure}

Fig. \ref{fig:NumericalSensitivity} presents plots of the CRBs for $\sigma_d$, as a function of the source separation $d/w_0$. In the Supp. Mat. we show similar plots associated with the parameters $c$ and $p$. 
In solid line we show the mean value of $\sigma_{d,\mathrm{CRB}}$ and, to illustrate the dependence of the sensitivity on the true values of the scene, we show a shaded area containing $90\%$ of the computed values of the CRB. The dual-MPLC configuration reduces the dependence on the three parameters, and also reduces the average value of the CRB compared to the one computed for the single-MPLC configuration. For the case of indistinguishable sources we included the averaged value of the QCRB, computed as in Ref. \cite{rehacek2017multiparameter}, we found a minimum difference of a factor of $\sim 4$, between the classical sensitivity of the 2-MPLC configuration and the QCRB.

\section{Experimental setup}

\begin{figure}[b]
\includegraphics[width=1\linewidth]{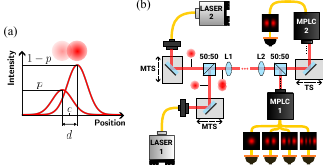}
\caption{\label{fig:scheme}
(a) The 1D profile of two Gaussian beams with the same waist $w_0$, separated by a distance $d$ and centered at $c$ (with respect to the first MPLC origin). The brightness imbalance is characterized by the parameter $p$, so that the triplet $\boldsymbol{\theta} = (d, c, p)^\top$ fully defines the scene. 
(b) Conceptual scheme of the setup. The left part corresponds to source generation, where motorized translation stages (MTS) are used to tune the positions of the two sources and change the scene. The brightness imbalance is controlled by adjusting the power in each source path via attenuators. The light is propagated through a telescope for mode-matching and then detected by the two demultiplexers, with one deliberately transversely shifted relative to the other (right part).}
\end{figure}

The experiment, Fig.~\ref {fig:scheme}(b), is performed using two independent, fiber-coupled continuous-wave (CW) lasers at \SI{1550}{\nano\meter} to generate two incoherent light sources. Each guided mode is passed through an optical isolator, a polarization controller, and a fibered attenuator, allowing for independent tuning of polarization and optical power. Once collimated into free space, they form Gaussian beams with a waist of $w_1\simeq \SI{1135}{\micro\meter}$. Each beam is reflected by a mirror mounted on a motorized translation stage (MTS) with a minimum step size of 20 nm, enabling fine adjustment of transverse displacement and precise control over spatial separation at the beam splitter. The combined beams are projected through a Glan-Thomson polarizer to ensure a common linear polarization, so that the obtained scene closely reproduces the image of two indistinguishable point light sources by an optical system with a Gaussian PSF. A telescope images and mode-matches the beams for optimal coupling into a demultiplexer (multiplane light converter, MPLC, PROTEUS-C from Cailabs) with waist $w_0 \simeq \SI{320}{\micro\meter}$ \cite{Morizur:10}. 
It effectively implements a unitary transformation from these modes to spatially separated Gaussian modes, each coupled into a distinct single-mode fiber for independent detection.
The light is split so that half of it is sent to another MPLC, transversely shifted with respect to the first one.

We limit ourselves to the one-dimensional problem, as the two components of the separation vector, $\boldsymbol{d} = (d_x, d_y)$, can be treated independently: each affects different sets of modes (horizontal versus vertical). For the estimations, we use the first four horizontal modes of the first MPLC, $\mathrm{HG}_{00}$, $\mathrm{HG}_{10}$, $\mathrm{HG}_{20}$, and $\mathrm{HG}_{30}$, along with the $\mathrm{HG}_{10}$ mode of the second MPLC (also referred to as $\mathrm{HG}_{10}^{\rm 2nd}$).
The corresponding intensities are measured simultaneously with photodiodes, while an additional detector records the total power of the scene.

\section{Multiparameter Estimations}
\subsection{Pre-procedure: Alignment, Mode-Matching, and Calibration}

The experimental procedure begins with mode-matching the two sources to the first demultiplexer by adjusting the telescope lens ratios to minimize the MPLC outputs corresponding to the $\mathrm{HG}_{20}$ and $\mathrm{HG}_{02}$ modes, which are sensitive to deviations of the source waist from the demultiplexer’s intrinsic waist; this adjustment is done once and then kept fixed. 
Coarse alignment of both sources is achieved by maximizing the $\mathrm{HG}_{00}$ mode, while the $\mathrm{HG}_{10}$ and $\mathrm{HG}_{01}$ modes, highly sensitive to tilt and misalignment relative to the demultiplexer origin, are minimized for fine tuning.

Being a twin of the first MPLC with identical waist $w_0$, the second device operates under the same telescope and mode-matching conditions optimized for the first, and is aligned to the sources using the same strategy with $\mathrm{HG}_{00}^{\mathrm{2nd}}$, $\mathrm{HG}_{10}^{\mathrm{2nd}}$ and $\mathrm{HG}_{01}^{\mathrm{2nd}}$. It is then transversely shifted by an amount $x_s=0.3w_0$ using a linear translation stage, so that part of the light is measured in a different mode basis.

Calibration is performed by moving each source individually along the horizontal axis of the transverse plane using motorized stages, while recording all outputs at each position. This step characterizes the system’s actual response and captures experimental imperfections such as misalignment, mode-matching errors, crosstalk, and deviations from an ideal Gaussian spatial profile.
By normalizing this data with the total power and fitting it with a high-order polynomial, we obtain the fraction of intensity as a function of the source position: $I_{i,j}(x_j)$, where $i \in \{0,1,2,3,4\}$ denotes the mode and $j \in \{1,2\}$ denotes the source.
Note that, in theory, the two sources---being identical---should exhibit the same response. In practice, however, $I_{i,1} \neq I_{i,2}$; this difference is quantified by a visibility of $99\%$ between the two beams.
When both sources are on, and since the light is incoherent, the fraction of intensity in mode $i$ is given by the sum of the contributions from the two sources:
    \begin{align}
\mu_{i}(\boldsymbol{\theta}) 
               &= p\,I_{i,1}\Big(c - \frac{d}{2}\Big) + (1-p)\,I_{i,2}\Big(c + \frac{d}{2}\Big).
    \end{align}

Here $\boldsymbol{\theta}=(d,c,p)^{\top}$ denotes the triplet of parameters that fully defines a scene (see Fig.~\ref{fig:scheme}(a)).
We explore sub-Rayleigh separations and centroids, and three different imbalances --  from equally bright sources to nearly ten times brighter -- as summarized in Table~\ref{tab:scan_params}.
The range of parameters is chosen to demonstrate that, with one experimental configuration, we can achieve simultaneous estimation of all three parameters with high accuracy and sensitivity. All experiments are performed in a relatively high photon flux regime (tens of micro-watts). Note that given the linearity of the device, it would be straightforward to extend the results to lower flux levels, with lower sensitivity as the photon flux is much lower, thus this configuration is much less sensitive to experimental imperfections than we can assess here at high flux.

\begin{table}
\caption{Ranges of parameters in the measurements. Each scene corresponds to one combination of the points listed for each parameter. Note that negative distances are used for the two slightly different sources as it produces a different scene.}
\label{tab:scan_params}
\setlength{\tabcolsep}{4pt} 
\begin{ruledtabular}
\begin{tabular}{lccc}
\multicolumn{4}{c}{\textbf{Distinguishable sources (99\% overlap)}} \\
\hline
Parameter & Range & Step & Number of points \\
\hline
$d$ & $[-0.15, 0.15]w_0$ & $\approx 0.03 w_0$ & 10\\
$c$ & $[-0.07, 0.07]w_0$ & $\approx 0.03 w_0$ & 5 \\
$p$ & $\{0.1, 0.3, 0.5\}$ & -- & 3 \\
\\
\multicolumn{4}{c}{\textbf{Indistinguishable sources}} \\
\hline
Parameter & Range & Step & Number of points \\
\hline
$d$ & $[0.05,0.3]w_0$ & $\approx 0.06 w_0$ & 5\\
$c$ & $[-0.07, 0.07]w_0$ & $\approx 0.03 w_0$ & 5 \\
$p$ & $\{0.1,0.3,0.5\}$ & -- & 3 \\
\end{tabular}
\end{ruledtabular}
\end{table}

\subsection{Results with distinguishable sources}

We generate measurement data by scanning the horizontal axis of the transverse plane with the two motors and varying the source powers (keeping the total flux fixed at approximately \SI{300}{\micro\watt}).

In total, $10 \times 5 \times 3 = 150$ distinct scenes are recorded. For each scene, the reference parameters are first determined by measuring each source individually using a quadrant detector and an auxiliary photodiode to record its position and power.
Next, both sources are measured simultaneously for \SI{0.01}{\second} at a \SI{10}{\kilo\hertz} sampling rate, producing 100 temporal measurement points (time bins) per scene; these data are used for parameter estimation. 

For a given time bin, we estimate our parameters from the measured fraction of intensities in all modes
$\mathbf{y} = (y_0, y_1, \dots, y_4)^\top$, using a maximum likelihood estimator under the assumption of multivariate Gaussian noise, 
by minimizing the following loss function:

\begin{equation}
\begin{aligned}
    \ell(\boldsymbol{\theta}) &= (\mathbf{y} - \boldsymbol{\mu}_{\boldsymbol{\theta}})^\top \, \boldsymbol{\Gamma}^{-1}\, (\mathbf{y} - \boldsymbol{\mu}_{\boldsymbol{\theta}})
\end{aligned}
\end{equation}

where $\boldsymbol{\Gamma}$ is the covariance matrix of the observation $\mathbf{y}$, and $\boldsymbol{\mu}_{\boldsymbol{\theta}}=(\mu_0(\boldsymbol{\theta}), \mu_1(\boldsymbol{\theta}),\dots, \mu_4(\boldsymbol{\theta}))^\top$ are the expected fractions of intensities for a given $\boldsymbol{\theta}$. We estimate $\boldsymbol{\Gamma}$ from the ensemble of 100 time bins and then treat it as known when analyzing a single bin. The minimization of the above cost function with respect to $\boldsymbol{\theta}$ is performed numerically using differential optimization scheme.

\begin{figure}
\includegraphics[width=1\linewidth]{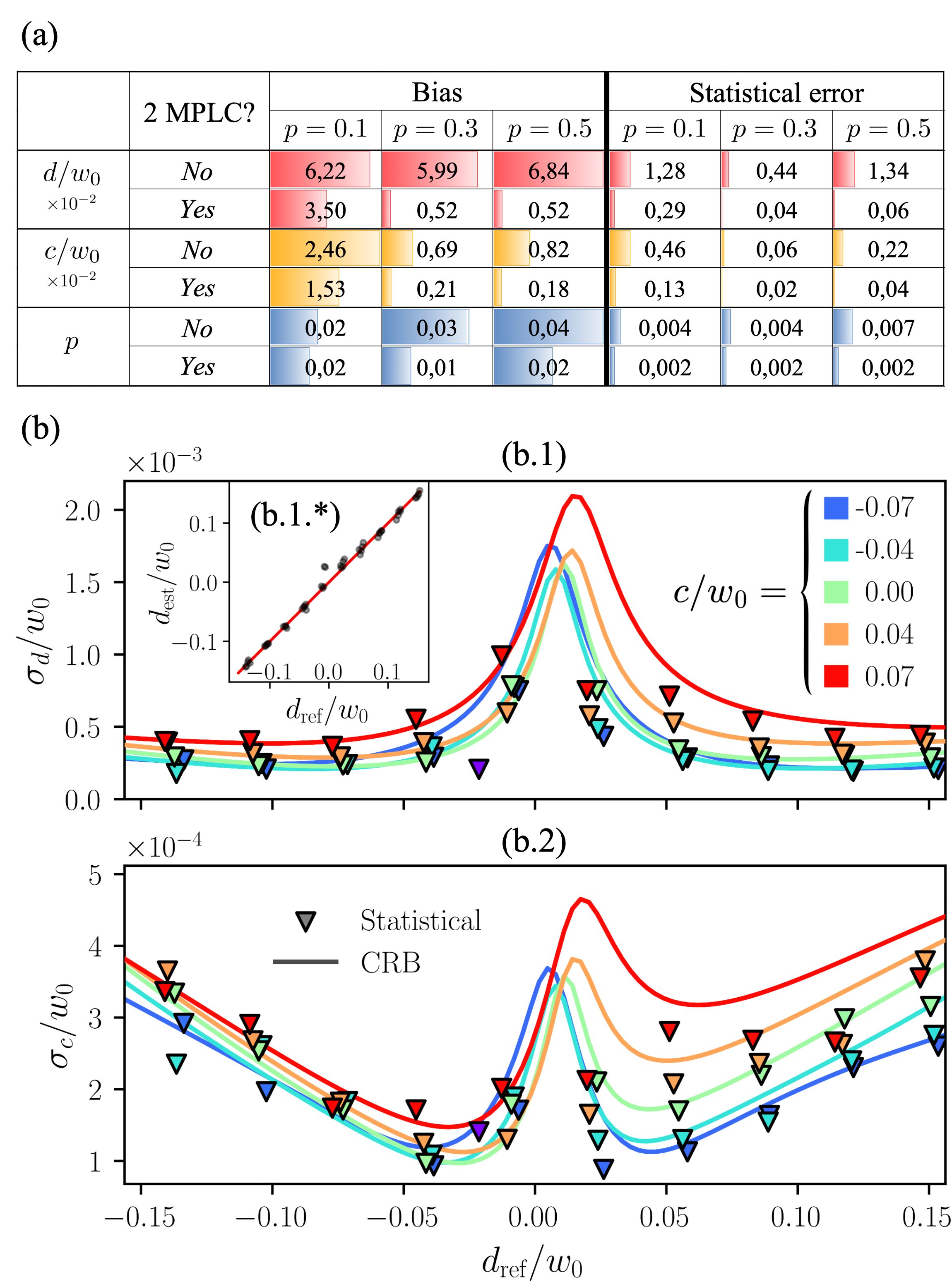}
\caption{\label{fig:exp_results}
Multiparameter estimation results on experimental data. Photon flux of $N = 10^{11}$ over \SI{0.1}{\milli\second}.
(a) Parameter-wise bias and sensitivity for $d$, $c$, and $p$. For each scene, the bias and sensitivity are computed from the ensemble of time-bin estimates, then averaged over all scenes. Results are grouped by brightness imbalance, and the table also specifies whether the dual-MPLC configuration was used.  
(b) Detailed results for the case $p = 0.3$.  
(b.1) Statistical errors $\sigma_d$ and (b.2) $\sigma_c$, obtained for each scene from the variability of the time-bin estimates of the separation $d$ and centroid $c$. Errors are shown as functions of the reference parameters $d_{\mathrm{ref}}$ and $c_{\mathrm{ref}}$ (see colors legend). Solid lines represent the Cramér–-Rao bounds computed from the experimental noise.  
(b.1.*) Mean estimate of separation $d_{\mathrm{est}}$ versus reference distances $d_{\mathrm{ref}}$. The red line denotes unbiased estimation.}
\end{figure}

Since the estimation is performed for each time bin, we have access to statistical errors $\sigma_{\alpha}$ which we compare to the diagonal elements of the inverse FI matrix $\sqrt{ \big(\mathcal{F}^{-1}\big)_{\alpha\alpha} }, \quad \alpha \in \{d, c, p\}$. We summarize the overall results in Fig.~\ref{fig:exp_results}. Panel~(a) provides the global results in tabular form, while panel~(b) presents the results for the ensemble of scenes with $p_{\mathrm{ref}} = 0.3$, along with a comparison to the corresponding CRB.

Our experiment demonstrates highly accurate and sensitive multiparameter estimation across a wide range of scene configurations, demonstrating respectively 2 to 4 orders of magnitude below the Rayleigh limit performances, enabled by the dual-MPLC setup. More specifically, in the dual-MPLC configuration, the achieved bias on the distance (see Fig.~\ref{fig:exp_results}(b.1.*)) and centroid is on the order of one hundredth of the beam waist , while the intensity imbalance is estimated with a one percent precision. The sensitivity on the separation measurement, quantified as the statistical error, reaches approximately $10^{-3}$ of the waist. As demonstrated in Fig.~\ref{fig:exp_results}(b.1) and (b.2), the statistical errors closely approach the Cramér–-Rao bounds computed from the experimental noise, both for separation and centroid estimation, demonstrating the efficiency of our estimator.

In contrast, performances in terms of bias deteriorate significantly when the second MPLC is omitted, primarily due to an unexpected degeneracy between $-d$ and $d$. In terms of sensitivity, there is also nearly an order of magnitude loss for both $d$ and $c$. These results demonstrate the capacity of our approach to perform accurate and precise multiparameter estimation without prior knowledge of the scene configuration, and highlight the crucial role of the second MPLC in resolving ambiguities and increase precision. 

Importantly the precision achieved is orders of magnitude below the diffraction limit, even though we can benefit from the slightly different sources with 99\% visibility. In the next section we explore the ideal case of indistinguishable sources to asses the performance of our setup in this scenario. 

\subsection{Results with indistinguishable sources} 

\begin{figure}
\includegraphics[width=1\linewidth]{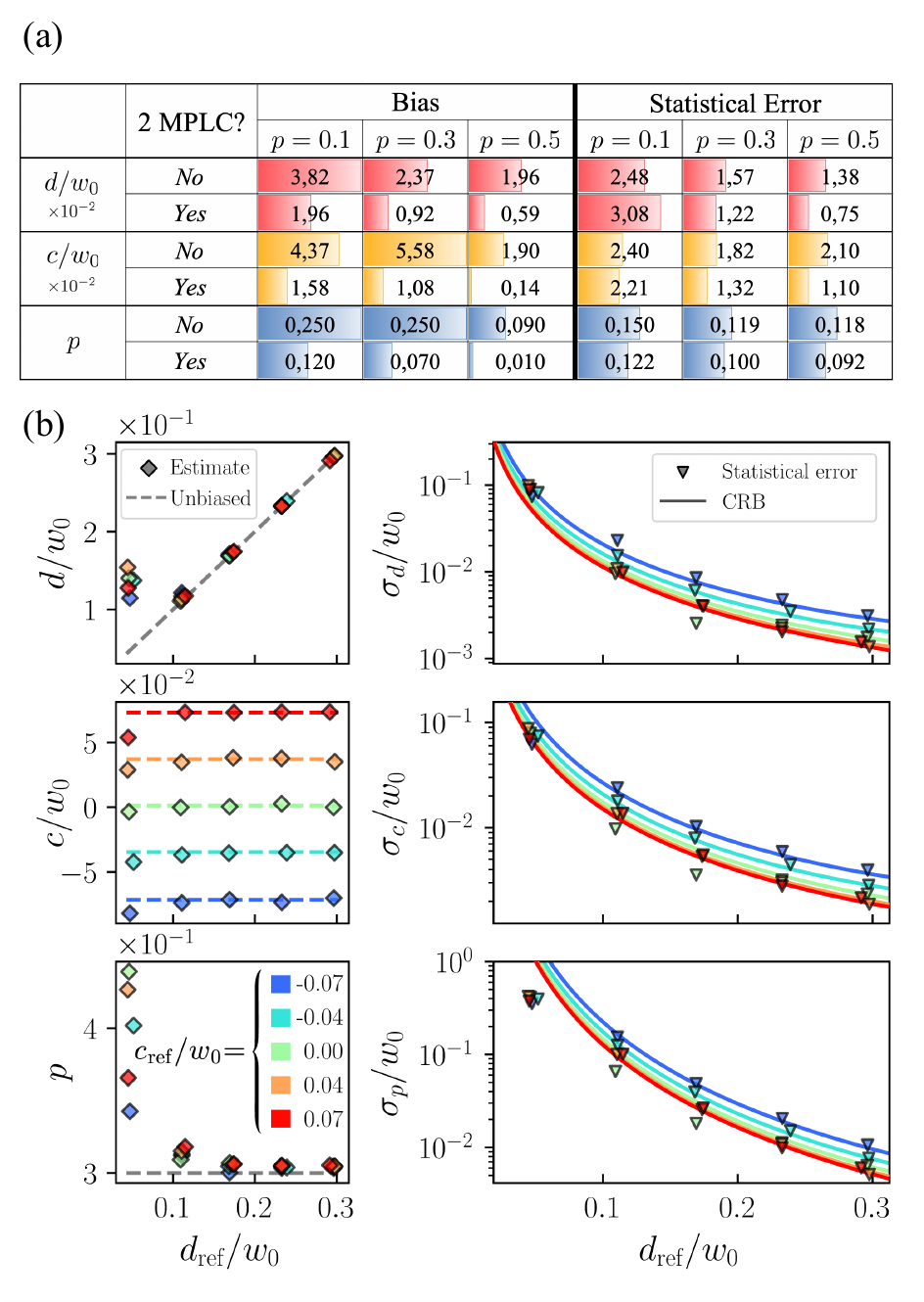}
\caption{\label{fig:results_simulated}
Estimation results on simulated data for the identical-source case. Photon flux of $N = 10^{11}$ over \SI{0.1}{\milli\second}.
(a) Same layout as in Fig.~\ref{fig:exp_results}(a).
(b) Specific case of $p_\mathrm{ref}=0.3$.
Left column: Mean parameter estimates (averaged over all time bins) as functions of the reference parameters $d_{\mathrm{ref}}$ and $c_{\mathrm{ref}}$ (see legend). Dashed lines indicate unbiased estimations.
Right column: Associated statistical errors, also shown as functions of the reference parameters $d_{\mathrm{ref}}$ and $c_{\mathrm{ref}}$.
}
\end{figure}

Depending on practical applications, it may be necessary to resolve two truly indistinguishable sources, in particular when no prior information is available about the sources. To evaluate how the setup would perform under such conditions, we generate data based on the calibration data from a single source. By summing the signals corresponding to two different positions, we reconstruct scenes similar to those analyzed previously, while ensuring source indistinguishability (see supplementary materials). The emulated scenes have separations up to $0.3w_0$ and only positive $d$ values, as the problem becomes rigorously symmetrical. The same estimation procedures are then applied, and the results are presented in Figure~\ref{fig:results_simulated}. 

It is remarkable that even in this scenario of indistinguishable sources, our dual-MPLC setup enables accurate and precise multiparameter estimation across a wide range of scenes, demonstrating the versatility and applicability of our approach. In terms of bias, performances are comparable to the previous case with slightly different sources, with biases on $d$ and $c$ remaining in the $10^{-2}w_0$ range. One can see that the dual-MPLC configuration significantly reduces the bias of the estimator compared to the single-MPLC case, making the experiment more reliable, a feature that cannot be captured via the CRB analysis. 

The obtained sensitivities are degraded compared to the previous case, but still remain in the $10^{-2}w_0$ range for $d$ and $c$ over the full range of parameters. From Fig.~\ref{fig:results_simulated}(b) one can see the efficiency of our estimator for a specific imbalance $p=0.3$. It can be observed that for separations below $0.1w_0$, estimators diverge significantly, both in terms of bias and sensitivity, but $d$ and $c$ can still be approximated up to $10^{-1}w_0$ even if $p$ cannot. Those results are consistent with the diverging CRB for small separation of identical sources. For separation greater than $0.1w_0$, the statistical error in separation $d$ and centroid $c$ drops below $10^{-2}w_0$.

\section{Conclusion}
We have experimentally demonstrated a simultaneous estimation of multiple parameters in the optical domain. In particular, we estimated the separation, centroid, and intensity imbalance for two incoherent point sources using a dual-SPADE architecture in the sub-Rayleigh regime, thereby obtaining superresolution without prior knowledge of the sources. We investigated two complementary classes of scenes: a realistic experimental regime with slightly distinguishable sources (99\% overlap), where the highest sensitivities are achieved, with separation uncertainties reaching $10^{-3}w_0$, and an idealized regime with indistinguishable sources, where separation and centroid can still be estimated at the $10^{-2}w_0$ level over a broad parameter range. By jointly addressing these two regimes, our results clarify the practical scope of multiparameter SPADE measurements for superresolution imaging. Although the experiment was performed in a high–photon-flux regime, the approach can be extended to lower-flux conditions with appropriate detection schemes, bringing it closer to some practical applications such as astronomical imaging. Furthermore, the method is not restricted to two sources and can be straightforwardly generalized to three or more incoherent emitters, with the addition of sources primarily increasing the number of parameters to be jointly estimated.

\begin{acknowledgments}

This project has received funding from the European Defence Fund (EDF) under grant agreement 101103417 EDF-2021-DIS-RDIS-ADEQUADE. Funded by the European Union. Views and opinions expressed are however those of the author(s) only and do not necessarily reflect those of the European Union or the European Commission. Neither the European Union nor the granting authority can be held responsible for them.
\end{acknowledgments}


\bibliography{references}

\end{document}


\title{Supplementary material - Multiparameter estimation for the superresolution of two incoherent sources}

\author{Antonin Grateau}
\author{Alexander Boeschoten}
\author{Tanguy Favin-Lévêque}
\author{Isael Herrera}
\author{Nicolas Treps}
\email{nicolas.treps@lkb.upmc.fr}

\affiliation{Laboratoire Kastler Brossel, Sorbonne Université, ENS-Université PSL, CNRS, Collège de France}
\maketitle

\section{Essential Material}

\begin{table}[h!]
\centering
\begin{tabular}{ll}
\hline
\textbf{Purpose} & \textbf{Component} \\
\hline
Lasers & SFL1550P laser diode (Thorlabs) -  Koheras ADJUSTIK HP E15 laser (NKT Photonics) \\
\cline{1-2}
Beam position control & Two V522 translation stages (Physik Instrumente, PI) for transverse displacement \\
\cline{1-2}
Demultiplexing & Two PROTEUS-C multiplane light converters (Cailabs), 10 modes each: \\
 & $\mathrm{HG}_{00}$, $\mathrm{HG}_{10}$, $\mathrm{HG}_{20}$, $\mathrm{HG}_{30}$, $\mathrm{HG}_{11}$, $\mathrm{HG}_{01}$, $\mathrm{HG}_{02}$, $\mathrm{HG}_{30}$, $\mathrm{HG}_{21}$, $\mathrm{HG}_{12}$ \\
\cline{1-2}
Detection & Five PDA50B2 photodiodes (Thorlabs) - Two NewFocus 2011 photodiodes (Newport) \\
\cline{1-2}
Position monitoring & PDQ30C quadrant detector (Thorlabs) \\
\cline{1-2}
Acquisition & USB-6366 (National Instruments) acquisition card\\
\hline
\end{tabular}
\label{tab:essential_material}
\end{table}

\section{Certifying positions and parameters}
\label{sec:certification}

We employ a quadrant detector (PDQ30C, Thorlabs) as a reference tool for position measurements. This device approaches the quantum limit for estimating the position of a single beam, although it cannot directly measure the separation between two simultaneously active beams.\\


During the measurement stage, the reference parameter $\boldsymbol{\theta_{\mathrm{ref}}}=(d_\mathrm{ref},c_\mathrm{ref},p_\mathrm{ref})^\top$ for a scene with two beams is obtained by sequentially turning each beam on and off and measuring their positions and power using the quadrant detector and a photodiode. This procedure allows us to determine the individual beam positions — and equivalently, the centroids and separation — as well as the brightness imbalance of the investigated scenes.

\section{Methods}

\subsection{Basic data processing}
We acquire the voltage outputs of all modes ($\mathrm{HG}_{00}$, $\mathrm{HG}_{01}$, $\mathrm{HG}_{02}$, $\mathrm{HG}_{30}$, and $\mathrm{HG}_{01}^{\mathrm{2nd}}$) at a sampling rate of \SI{10}{\kilo\hertz}. The voltages are then converted to optical powers after correction for background offsets and application of the gain factors. 
The resulting time series are normalized sample-wise by the total scene power, measured using an additional photodetector, to yield the observation vector $\mathbf{y}(t) = (y_0(t), \dots, y_4(t))^\top$. This vector represents the fraction of optical power in each mode as a function of time and will serve as the input for maximum likelihood estimation of the model parameters.

\subsection{Calibration curves}
\begin{figure}[h]
\centering
\includegraphics[width=1\linewidth]{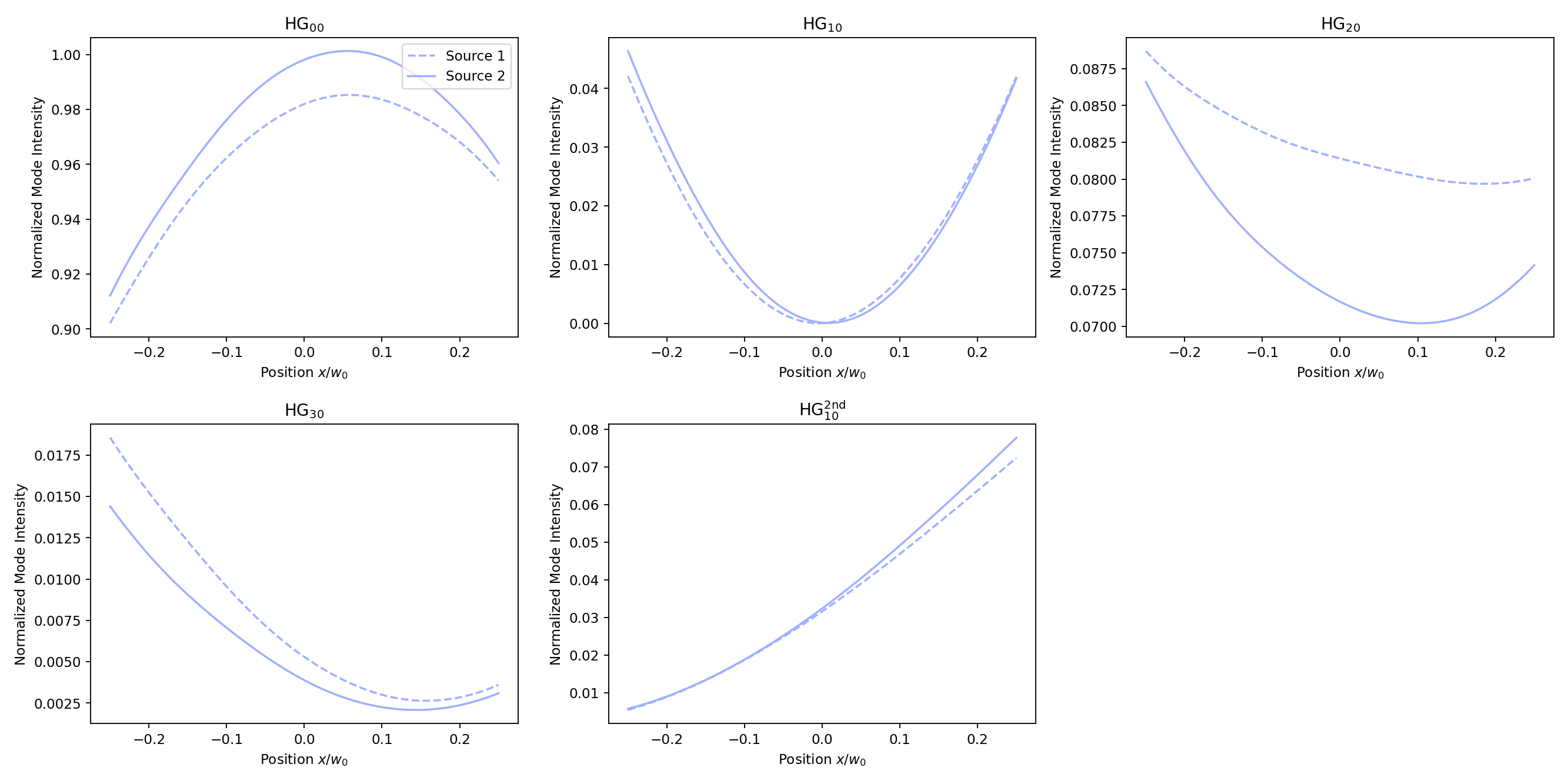}
\caption{Calibration results. Each source is horizontally displaced in the transverse plane. For each position in the range $[-0.3,0.3]w_0$, the output mode intensities are recorded. The plotted curves show the mode intensities normalized by the total measured power as a function of the source position, for each source.}
\label{fig::calibration}
\end{figure}

The demultiplexers are calibrated by individually translating each source along the horizontal axis of the transverse plane using motorized stages, while recording the output intensities of all modes at each position. This procedure characterizes the actual response of the system and accounts for experimental imperfections. The two sources do not exhibit perfectly identical calibration curves, as shown in Fig.~\ref{fig::calibration}. However, the difference is small for most of the channels, since the spatial overlap between the two sources is approximately $99\%$.

\subsection{Measurement sequence}

A measurement of a given scene consists of recording all mode output powers, along with the total power for normalization. Data are acquired at a sampling rate of \SI{10}{\kilo\hertz} while both sources are active, with each acquisition spanning \SI{0.01}{\second} (yielding 100 time samples). To obtain reference parameters $\boldsymbol{\theta}_{\mathrm{ref}} = (d_{\mathrm{ref}}, c_{\mathrm{ref}}, p_{\mathrm{ref}})^\top$, each source is also measured individually (as detailed in Section~\ref{sec:certification}).

A measurement sequence is the procedure used to acquire data over a set of distinct scenes. In our experiments, the parameters spanned the ranges $d \in [-0.15, 0.15]\,w_0$, $c \in [-0.07, 0.07]\,w_0$, and $p \in \{0.1, 0.3, 0.5\}$. The parameters $d$ and $c$ are adjusted automatically by translating the two sources in the transverse plane using motorized stages. In contrast, $p$ must be changed manually by tuning the power attenuators of each source, which requires opening the environmental enclosure. Because this procedure introduces additional drift on top of the natural temporal drift, we recalibrated the individual sources after setting each new value of $p$. Although the differences between these calibrations are subtle, they are critical for multiparameter estimation, as even modest systematic effects can cause the estimates to deviate significantly from the true values.

\subsection{Estimation}\label{sec::estimation}

We define the estimator $\check{\boldsymbol{\theta}}$ as the value that minimizes the weighted residual between the observation vector $\mathbf{y}$ (defined above) and the forward model $\boldsymbol{\mu}(\boldsymbol{\theta}) = (\mu_1(\boldsymbol{\theta}),\mu_2(\boldsymbol{\theta}), \dots)^\top $ defined in the main text (eq. (2)):
\begin{equation}
    \check{\boldsymbol{\theta}} = \arg \min_{\boldsymbol{\theta}}  \qquad(\mathbf{y} - \boldsymbol{\mu}_{\boldsymbol{\theta}})^\top \, \boldsymbol{\Gamma}^{-1}_{\boldsymbol{\theta}}\, (\mathbf{y} - \boldsymbol{\mu}_{\boldsymbol{\theta}})+ \log \det \boldsymbol{\Gamma_\theta}
    \label{eq::estimator}
\end{equation}

\begin{figure}[h!]
    \centering
    \includegraphics[width=0.5\linewidth]{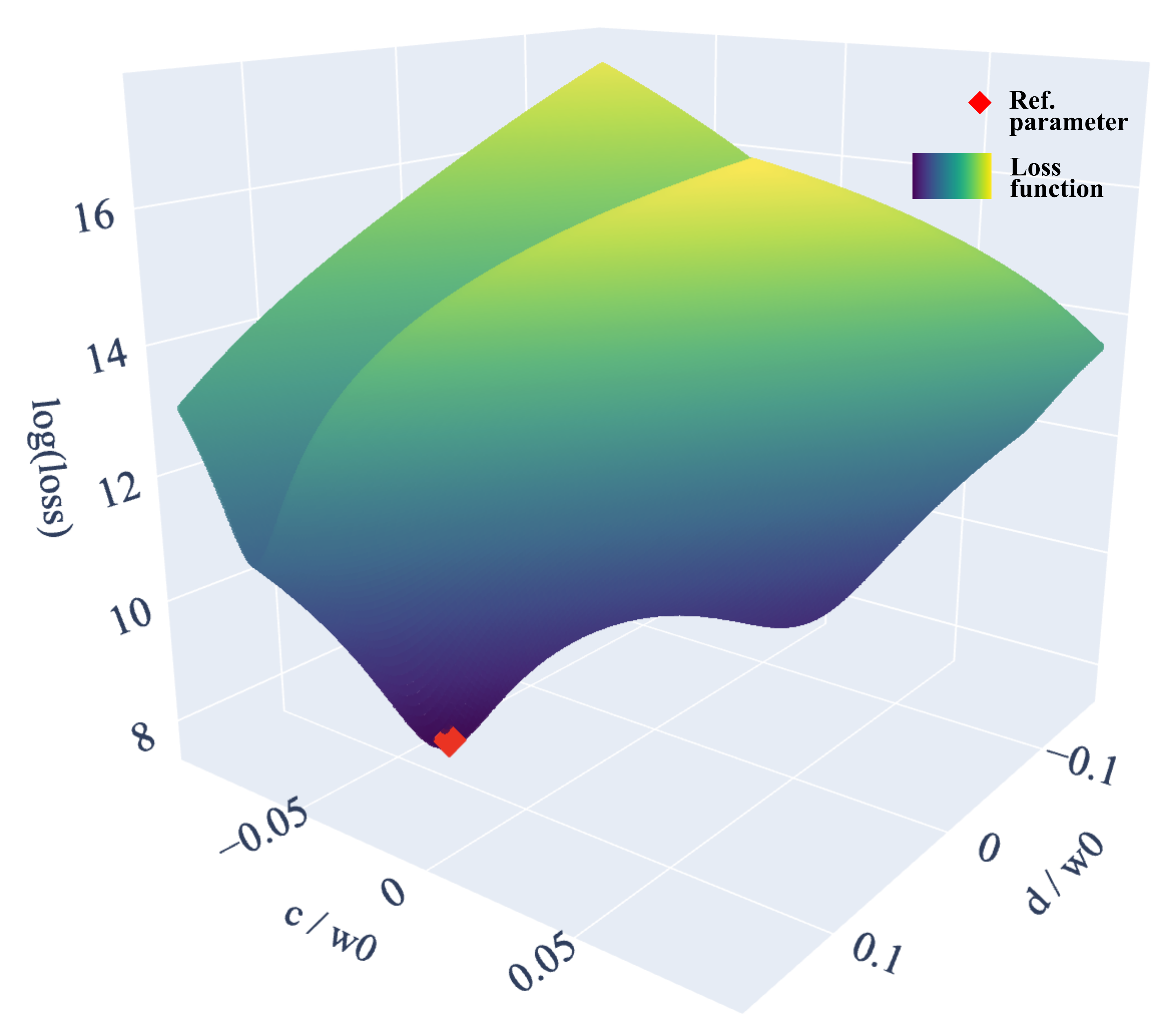}
    \caption{Loss function landscape in the search space. $d_\mathrm{ref}=28.3$, $c_\mathrm{ref}=-11.4$, $p_\mathrm{ref}=0.09$. For a given time bin observation, we plot the loss function as a function of $\theta$, keeping $p$ fixed at its reference value for visualization.  We observe that the minimum of the loss function occurs very close to the reference parameter.
}
    \label{fig:loss_function}
\end{figure}

This corresponds to a maximum likelihood estimate, assuming that the observations follow a normal distribution with a parameter-dependent mean $\boldsymbol{\mu}_\theta$ and covariance matrix $\boldsymbol{\Gamma_\theta}$. The covariance matrix is estimated from the 100 samples of the observation vector of each scene, and is treated as known during estimation for each time bin.
The optimal parameters are then determined using gradient-based optimization, which reliably identifies the global minimum of the loss function calculated over the full parameter space (see Fig.~\ref{fig:loss_function}).

\subsection{Emulated scenes with indistinguishable sources}

To simulate measurements of a scene featuring two identical sources, we utilize time series acquired during the calibration of a single source.  Consequently, we select two calibration points at positions $x_1$ and $x_2$
to achieve the target separation and centroid. Each signal is then decomposed into its mean value and a residual time series. The mean values are scaled to reflect the desired power imbalance, and both residual time series are summed, resulting in a conservative noise estimate where the electronic noise is effectively doubled (as our setup is electronic-noise limited).

\section{Accuracies of the estimates}
\subsection{Distinguishable sources (99\% overlap, experimental data)}
\begin{figure}
    \centering
    \includegraphics[width=0.65\linewidth]{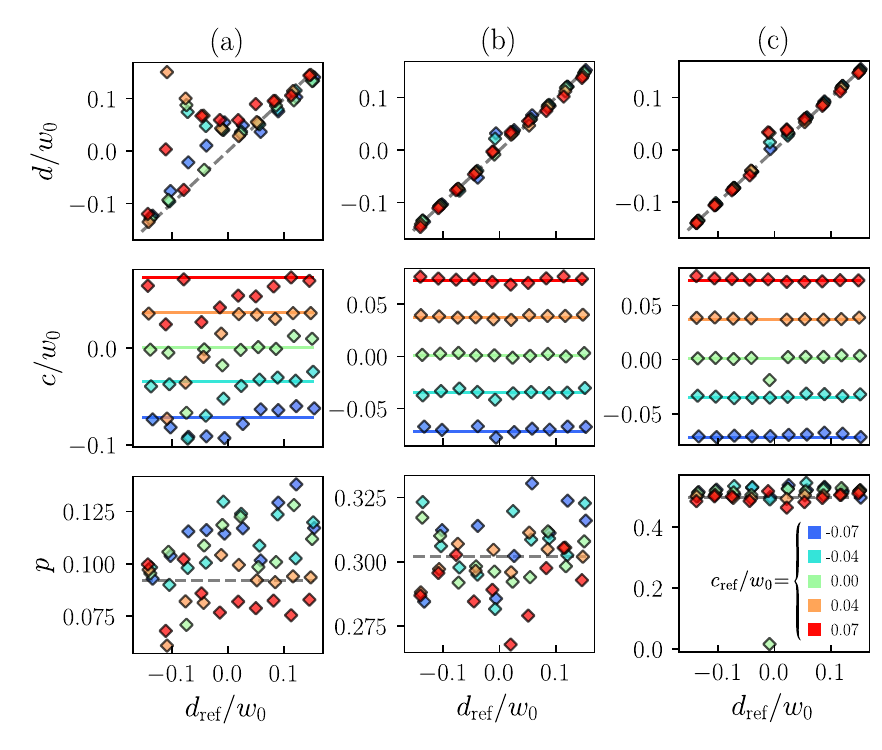}
    \caption{Estimated parameters with respect to reference separation for $p=0.1$ (a), $p=0.3$ (b) and $p=0.5$ (c), and for different centroids (see color maps). Dashed lines correspond to unbiased estimate.}
    \label{fig::bias_plot_all}
\end{figure}

See Fig.~\ref{fig::bias_plot_all}

\subsection{Indistinguishable sources (emulated data)}
\begin{figure}
    \centering
    \includegraphics[width=0.65\linewidth]{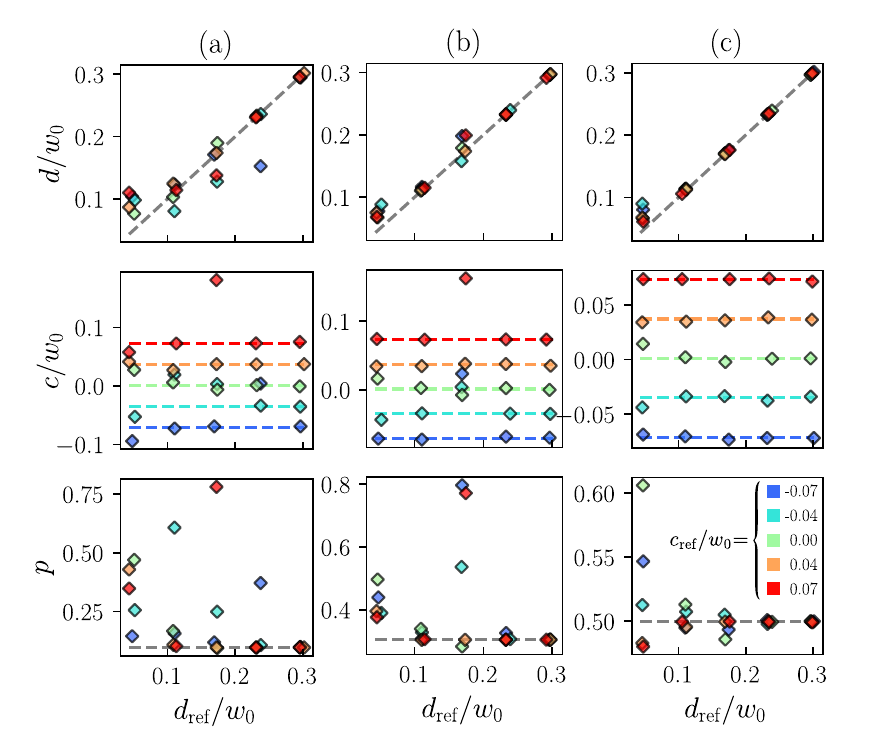}
    \caption{Estimated parameters with respect to reference separation for $p=0.1$ (a), $p=0.3$ (b) and $p=0.5$ (c) over emulated measurement of two indistinguishable sources for different centroids (see color maps). Dashed lines correspond to unbiased estimate.}
    \label{fig::bias_plot_all_simulated}
\end{figure}

See Fig.~\ref{fig::bias_plot_all_simulated}
\section{Fisher Information}


\subsection{For an ideal system}\label{eq::ideal-system-fim}
The Fisher information matrix of a system where the total number of photons is evenly distributed between two MPLCs, assuming only shot noise and absence of readout noise and background is given by


\begin{equation}
    \mathcal{F}_{\alpha,\beta}=\frac{N}{2}\sum_{i}\frac{1}{\mu_i}\frac{\partial\mu_{i} }{\partial \theta_\alpha}\frac{\partial\mu_{i} }{\partial \theta_\beta}
    \label{eq::FIM-ideal}
\end{equation}

The functions $\mu_{i}$ are defined in Eq. 2 of the main text. Each function  $\mu_{i}$ is associated to the $i$-th mode. In practice, we only use the first four horizontal modes of the first MPLC: $\mathrm{HG}_{00}$, $\mathrm{HG}_{10}$, $\mathrm{HG}_{20}$, and $\mathrm{HG}_{30}$, and the $\mathrm{HG}_{10}$ mode of the second MPLC. For both MPLCs, the mode  $\mathrm{HG}_{00}$ contains most of the detected photons, hence, when using the 2-MPLC configuration, almost half of the photons are not measured. The covariance matrix is bounded by the inverse of the FI matrix as $ \rm{Cov}_{\alpha,\beta}\geq (\mathrm{\mathcal{F}}^{-1})_{\alpha,\beta}$. 

To calculate the bounds shown in Fig. 1 of the main text, we compute the square root of the diagonal terms of the inverse of the FI matrix,  $\sigma_{\alpha}\geq\sqrt{(\mathcal{F}^{-1})_{\alpha,\alpha}}$.
To simulate the situation of distinguishable or undistinguishable sources we considered $I_{i,1}=I_{i,2}$ or $I_{i,1}\neq I_{i,2}$ respectively in the definition of $\mu_i$.   

\begin{figure}[h]
\includegraphics[width=1\linewidth]{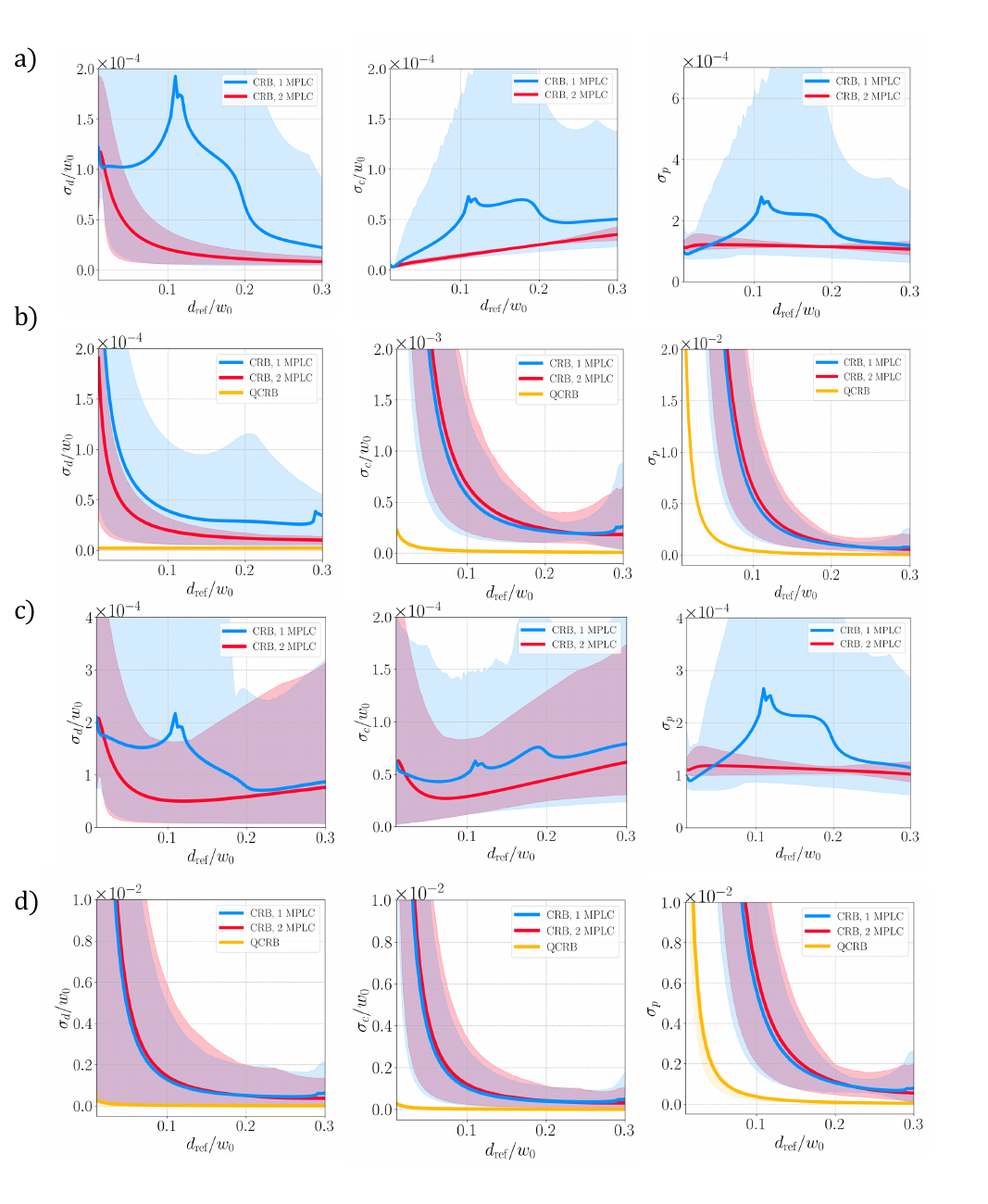}
\caption{\label{fig:NumericalSensitivitysupp} Cramér-Rao bounds for, from left to right,  $d$, $c$ and $p$. We show in blue/red the results for the case of 1/2 MPLC configuration associated with: a),c) distinguishable and b),d) undistinguishable sources. In solid lines we present the average precision on the parameter $\alpha$. The corresponding shaded area is the region containing $90\%$ of the computed values. For a)-b), we use an assemble of scenes defined by $d\in [0.01,0.3]w_0$, $c\in [-0.15,0.15]w_0$  and $p= 0.5$, with $N=10^{11}$ signal photons. For c)-d), we use an assemble of scenes defined by $d\in [0.01,0.3]w_0$, $c\in [-0.15,0.15]w_0$  and $p\in[0.05, 0.5]$, with $N=10^{11}$ signal photons. For distinguishable sources we also show the quantum CRB. %
}
\end{figure}

In Fig. \ref{fig:NumericalSensitivitysupp} we present the CRB associated with the three parameters $\{d,c,p\}$, similar to the ones shown in Fig. 1 of the main text. We observe that the 2-MPLC configuration has a significant impact in the precision of the three parameters for the case of distinguishable sources. 

\subsection{For the experiment}
The calculation of the Cramér-Rao Bound (CRB) presented in Fig.~3 of the main text follows the assumptions detailed in Section~\ref{sec::estimation}, where each channel is modeled by a Gaussian distribution. Consequently, the Fisher Information Matrix (FIM) is expressed as: \begin{equation} \mathcal{F}_{\alpha, \beta} = \left(\frac{\partial \boldsymbol{\mu}_{\boldsymbol{\theta}}}{\partial \theta_\alpha}\right)^\top \boldsymbol{\Gamma}_{\boldsymbol{\theta}}^{-1} \left(\frac{\partial \boldsymbol{\mu}_{\boldsymbol{\theta}}}{\partial \theta_\beta}\right) \end{equation} where the second-order term involving the derivative of the covariance matrix has been omitted, as $\boldsymbol{\Gamma_\theta}$
is slowly varying with respect to the parameters. The mapping between the true parameters and the noise covariance matrix $\boldsymbol{\Gamma_\theta}$
was established empirically across all measured scenes. Specifically, we performed a polynomial regression on the covariance entries measured across all experimental scenes, allowing for accurate interpolation of $\boldsymbol{\Gamma_\theta}$ for arbitrary centroids and separations at a fixed power imbalance.